# Commissioning and operation experience with the China ADS Injector-I CW linac

Fang Yan†, Huiping Geng, Cai Meng, Yaliang Zhao, Huafu Ouyang, Shilun Pei, Rong Liu, Feisi He, Tongming Huang, Rui Ge, Yanfeng Sui, Qiang Ye, Xiaoping Jing, Fengli Long, Jungang Li, Quanling Peng, Dizhou Guo, Zusheng Zhou, Haiyin Lin, Xinpeng Ma, Qunyao Wang, Guangwei Wang, Hua Shi, Gang Wu, Shengchang Wang, Ningchuang Zhou, Qiang Ma, Zhenghui Mi, Peng Sha, Xinying Zhang, Yaoyao Du, Jun He, Huizhou Ma, Lingda Yu, Ying Zhao, Xiaoyan Zhao, Fang Liu, Yanhua Lu, Lin Bian, liangrui Sun, Rui Ye, Xiaohua Peng, Dayong He, Ouzheng Xiao, Yao Gao, Zhenghua Hou, Yuan Chen, Xiangchen Yang, Hongyan Zhu, Bo Li, Lan Dong, Heng Li, Xitong Sun, Linglang Dong, Ping Su, Jianping Dai, Jianli Wang, Shaopeng Li, Jianshe Cao, Yunlong Chi, Weimin Pan

*Key Laboratory of Particle Acceleration Physics and Technology*

*Institute of High Energy Physics (IHEP), Chinese Academy of Science, Beijing 100049, China*

**Abstract:** The 10 MeV accelerator-driven subcritical system (ADS) Injector-I test stand at Institute of High Energy Physics (IHEP) is a testing facility dedicated to demonstrate one of the two injector design schemes [Injector Scheme-I, which works at 325 MHz], for the ADS project in China. The Injector adopted a four vane copper structure RFQ with output energy of 3.2 MeV and a superconducting (SC) section accommodating fourteen $\beta_g$=0.12 single spoke cavities, fourteen SC solenoids and fourteen cold BPMs. The ion source was installed since April of 2014, periods of commissioning are regularly scheduled between installation phases of the rest of the injector. Continuous wave (CW) beam was shooting through the injector and 10 MeV CW proton beam with average beam current around 2 mA was obtained recently. This contribution describe the results achieved so far and the difficulties encountered in CW commissioning.

## I. INTRODUCTION

ADS project in China was launched in year of 2011 intending to develop the concept and design of a 1.5 GeV high intensity SC linac with the aim of building a demonstration facility for accelerator-driven subcritical system (ADS) in multiple phases. The driver linac will be operating in continuous wave (CW) mode and delivering 15 MW beam power eventually. The linac includes two major sections: the injector section and the main linac section. The injectors accelerate the proton beams up to 10 MeV and the main linac boost the energy from 10 MeV up to 1.5 GeV.

To develop key technologies and skills of the design and operation for high intensity CW linac, and demonstrate the design schemes, two different injectors on basis of two different frequencies are proposed [1, 2], fabricated and commissioned in two different institutes independently. Institute of High Energy Physics (IHEP) is responsible for Scheme I (so called Injector-I) which is based on 325 MHz (the same frequency with the main linac) room temperature RFQ and SC spoke cavities of the same RF frequency and Institute of Modern Physics (IMP) is responsible for scheme II (so-called Injector II) based on 162.5 MHz room temperature RFQ and Half wave resonator (HWR) technology of the same frequency. Finally only one scheme will be chosen and two identical injectors will be built and operated as a hot spare stand-by.

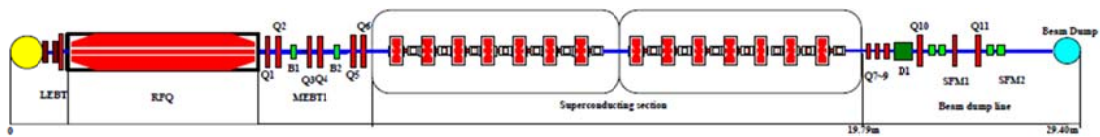

FIG. 1: The general layout of the ADS injector-I linac.

* Work supported by CAS Strategic Priority Research Program-Future Advanced Nuclear Fission Energy (Accelerator-Driven Sub-critical System) and National Natural Science Foundation of China, under contract NO. 11405190.

† yanfang@ihep.ac.cn

The specifications of the injector-I are listed in table I. The benefit for choosing 325 MHz scheme is that there is no frequency jump from the injector to the main linac. For the same design of main linac, the acceptance over input beam emittance ratio would be bigger for 325 MHz option than half frequency choice, if keeping space charge effect of the entrance beam to be the same for both injector schemes. Because the specifications of average beam current (10 mA) at the end of the driver linac is the same no matter which scheme is adopted for the injector. Otherwise the bunch intensity would be twice bigger for the half frequency choice if the entrance beam sizes keeping to be the same in both schemes. Decreased space charge effect with smaller bunch current is beneficial for controlling the beam qualities especially for low energy section. In the other hand, we do not want to decrease the acceptance also as the beam loss restriction is rigorous for high intensity CW machine, especially the high energy part. However if keeping the space charge effect of the entrance beam to be the same and in the mean while keeping the acceptance over input beam emittance ratio to be the same for both injector schemes, the cost of the main linac has to be increased to get bigger acceptance for the injector on basis of half frequency because the entrance beam size has be enlarged.

The designed injector output energy (10 MeV), which is also entrance energy of the main linac, is determined by the at least energy for the possibility of local compensation mechanism realization for the main linac because the injector is designed to be backed up by another one as mentioned before.

The general layout of the ADS injector-I linac is shown in figure 1. the Injector-I is composed of an Electron Cyclotron Resonance (ECR) ion source, a Low Energy Beam Transport (LEBT) line, a 4-vane type copper structure Radio Frequency Quadrupole (RFQ), a Medium Energy Beam Transport (MEBT) line, a superconducting section adopting single spoke cavity for acceleration, an energy divergence system and a beam dump line.

One of the main difficulties of Injector-I CW operation is the power dissipation of the CW room temperature RFQ, another one is the design and operation of the single spoke SC cavity with β geometry of 0.12 which is the smallest beta developed ever in the world for this kind of cavities. It is not only the first try of the spoke cavity with such small beta, but also the first spoke type cavity verified by beam until now, let alone CW beam. The cavity beta is selected on basis of a tradeoff between the RFQ output energy and the downstream SC cavity accelerating efficiency. Higher RFQ output energy implies higher power needed for feeding into the cavity and thus higher heat deposit on the cavity surface. Lower entrance energy of the SC section needs even smaller geometry beta of cavity for getting higher cavity efficiency. However the ratio of peak field over accelerating gradient is actually bigger with further smaller beta of the cavities because of even narrower accelerating gap.

Table I: ADS Injector-I test facility specifications

| ADS Injector-I test facility specifications | |
|---|---|
| Particle | Proton |
| RF frequency (MHz) | 325 |
| Output Energy (MeV) | 10 |
| Average Current (mA) | 10 |
| Beam power (kW) | 100 |
| Duty factor (%) | 100 |

The detailed physics design and design philosophy of the Injector-I are presented in reference [3-4]. The injector is designed to accelerate the proton CW beam up to 10 MeV with average beam current of 10 mA, we achieved maximum 2.1 mA CW proton beam with energy of 10 MeV at the exit of the Injector-I linac recently. This paper will focuses on the infrastructure fabricated and the experimental results obtained during the different stage beam commissioning of ADS injector-I test facility since April of 2014 and the challenging we encountered for pursuing the final goal.

## II. COMMISSIONING OF THE INJECTOR-I

### A. Ion source and LEBT

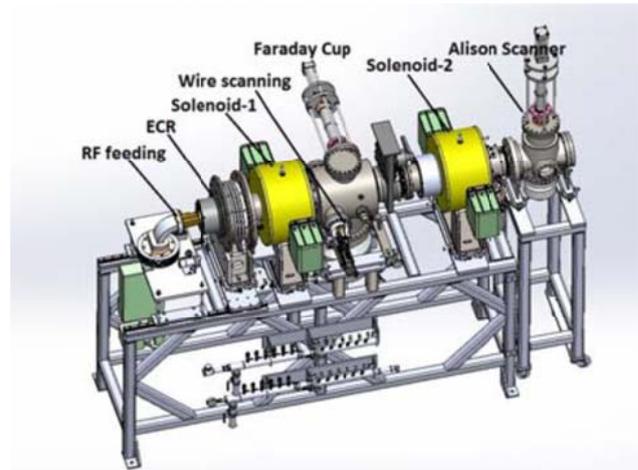

FIG. 2: Off-line layout of 2.45GHz ECR proton source and LEBT [5].

Detail design of the ion source and LEBT can be found in reference [5-6]. The schematic diagram of the ECR source and LEBT system are shown in Fig. 2. The source could operate in CW or pulsed mode providing with 35 keV CW or pulsed proton beam ($H^+$). The LEBT chopper located at the entrance of RFQ chopping the beam with pulse width starting from 20 μs or even less with repetition frequency of 1 Hz up to 50 Hz. The rise and down time of the chopper, as shown in figure 3, is smaller than 20 ns as measured at the MEBT FCT downstream RFQ. The ACCT1 locating between the chopping system and RFQ is used for the beam current measurement. Maximum average/peak current of 13 mA could be delivered at the entrance of RFQ.

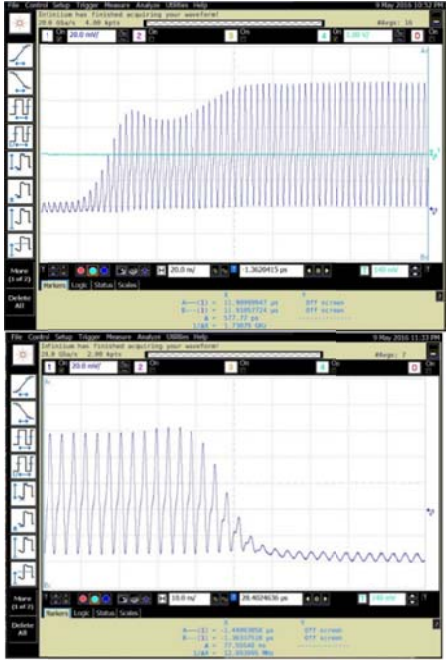

FIG. 3: LEBT chopper rise time (upper trace) and fall time (lower trace) measured at the MEBT FCT.

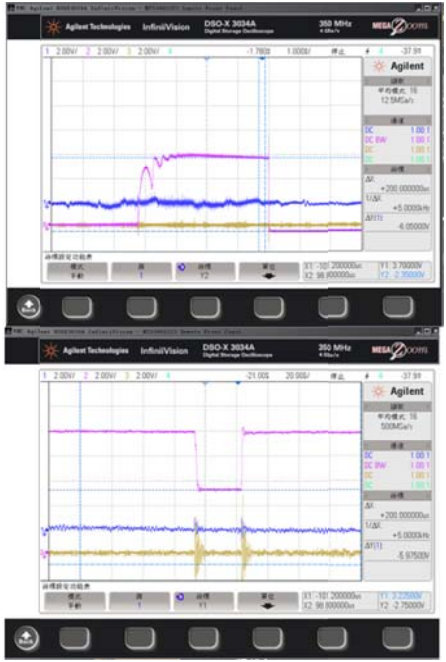

FIG. 4: ACCT1 signal with upstream chopper and downstream RFQ off (upper graph), ACCT1 signal with chopper on and RFQ power off (lower graph).

It is pretty tricky for the RFQ beam transmission measurement. As shown in figure 4 (upper), the current signal of ACCT has a slope from the front to the end of the pulse and the baseline in the front and at the end is not in the same level. The measurement has to be carried out according to the beam signal at the very end of the pulse with the baseline behind as shown also in the upper graph of figure 4. Besides, the chopper high voltage has an effect on the ACCT signals. As shown in the lower graph of figure 4, when the chopper is turned on, the part of beam pulse being chopped still appears in ACCT1 but with higher readings, while the beam signal going through the chopper gives lower reading comparing with chopper off. The RFQ power is similar, but has a smaller effect. If taking the beam signal with chopper on and RFQ off for the beam transmission measurement, the results would be ~1% higher than turning all of them off during the measurement.

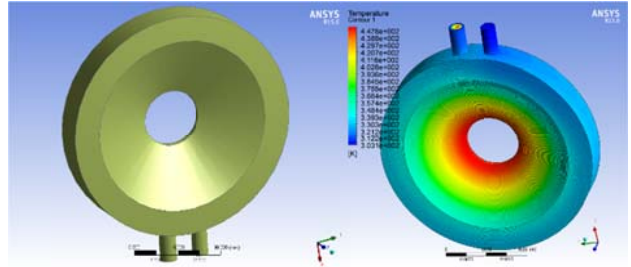

FIG. 5: Movable aperture of the LEBT section [7].

For producing lower beam current used for the initial CW commissioning, a moveable aperture, as shown in figure 5, was installed right after the first LEBT solenoid replacing the wire scanner showing in figure 2. However, the alignment of this device turns out to be a problem. It was considered to be the main reason of the lower transmission for the beam going through RFQ with smaller current, the detail will be covered in the next section.

Alison detector installed in a moveable diagnostic bench was used for the LEBT emittance measurement. Twiss parameters close to the designed values at the entrance of RFQ were achieved by adjusting the LEBT solenoid settings. 5% background was assumed. The simulation and measurement results reported in reference [4].

### B. The RFQ accelerator

The 325 MHz RFQ bunches and accelerates the 35 keV beam from the ion source to 3.2 MeV. The physics design of the 4.7 m long 4-vane type copper structure RFQ with π mode stabilizers is described in detail elsewhere [3]. The design parameters are summary in table II for easy reference. The RFQ is composed of two resonantly coupled physical segments and each segment includes two technical modules connected together with flanges. Totally four couplers are mounted on the RFQ and two couplers on one physical segment. 64 tuners are plugged symmetrically on four sections for field flatness tuning and frequency adjustment. Totally 80 main cooling channels are designed for the RFQ vane and wall heat dissipation. There are also auxiliary cooling channels for the inner and outer conductor of the feeding forward power couplers (FPC) and the plug tuners. Figure 6 shows one module of the ADS Injector-I RFQ, the left figure shows the cross section view of the RFQ and the side view is shown on the right figure.

The installation of the RFQ test stand was finished on May of 2014. Fig. 7 shows the RFQ test stand in the tunnel with one of the FPC mounted on the cavity as shown in the left figure of figure 7 and with all the assembling installed including the cooling channels as shown in the right figure. The layout of the RFQ test stand is shown in Figure 8. A movable diagnostic bench with two slits, one DCCT, two FCTs, four quadrupoles, four BPMs and a simple beam dump line was mounted right after the RFQ for the commissioning.

TABLE II. Main design parameters for the Injector scheme-I RFQ comparing with LEDA RFQ design [8-9].

| Parameters | Value |
|---|---|
| Structure type | 4 vane |
| Frequency (MHz) | 325 |
| Injection energy (keV) | 35 |
| Output energy (MeV) | 3.21 |
| Output peak current (mA) | >10 |
| RF duty factor | 100% |
| Inter-vane voltage $V$ (kV) | 55 |
| Designed beam transmission | 98.7% |
| Average bore radius $r_0$ (mm) | 2.775 |
| Maximum surface field (MV/m) | 28.88 (1.62Kilp.) |
| Cavity power dissipation (kW) | 272.94 [1.4* $P_{superfish}$ (194.96)] |
| Total power (kW) | 305.07 |
| Avg. Copper power/Length (kW/m) | 58.5 |
| Input norm. rms emittance(x/y/z)($\pi$ mm mrad) | 0.2/0.2/0 |
| Output norm. rms emittance(x/y/z) ($\pi$mm.mrad/$\pi$MeV-deg) | 0.21/0.21/0.058 |
| Vane length (cm) | 467.75 |
| Accelerator length (cm) | 469.95 |
| Cross dimension (mm× mm) | 260×260 |
| Gap1(entrance) (cm) / Gap2(exit) (cm) | 1.10 / 1.10 |

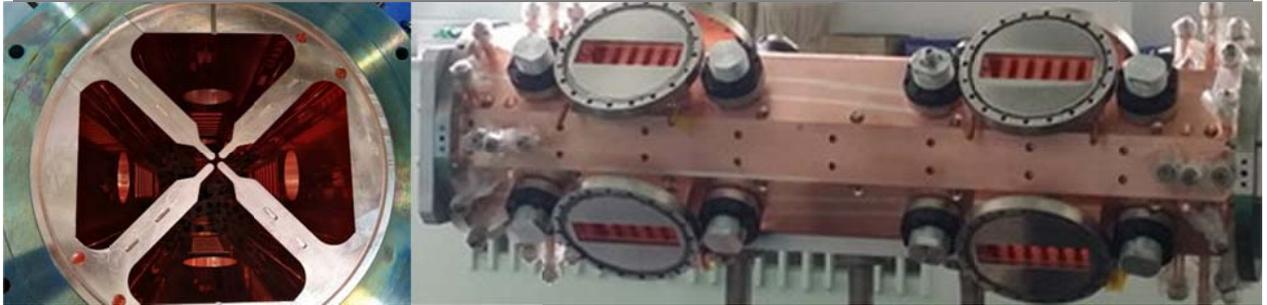

FIG. 6: One module of the ADS Injector-I RFQ: cross section view (left), side view (right).

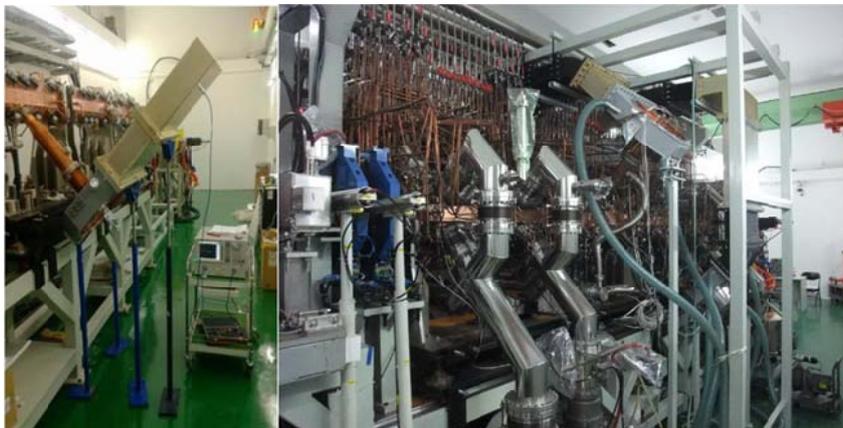

FIG. 7: The RFQ test stand in the tunnel with one of the feeding forward power coupler mounted on the cavity (left) and with all the assembling installed including the cooling channels (right).

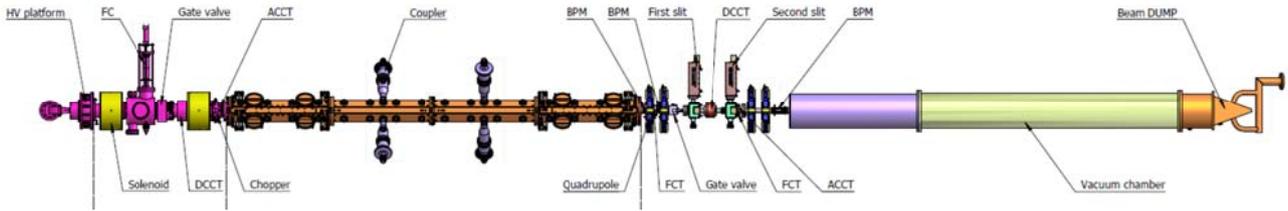

FIG. 8: The RFQ test stand layout.

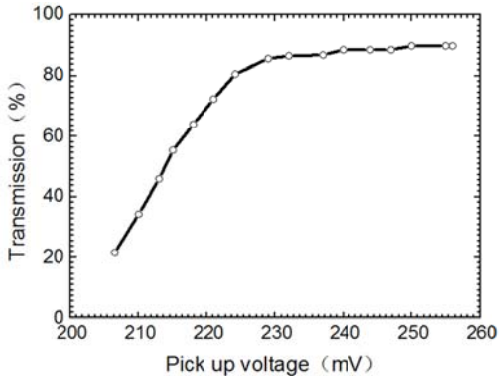

FIG. 9: Verification of tank level with transmission scan.

The RF commissioning starts from May 15[th] of year 2014, beginning with short pulse of 100 μs and 1% RF duty factor, gradually ramping the power to designed value. Then the pulse width was progressively expended up to 1 ms with full power until 50% duty factor being achieved. After reaching over 10 ms, the RF duty factor was increased by raising the repetition frequency while the pulse width being fixed. In the meanwhile CW conditioning was processed alternately. After about 2 months conditioning, 99.97% RF duty factor was reached with pulse width of 12.5 ms, repetition frequency of 79.975 Hz and 250 kW in cavity power. While switching to CW mode, the maximum in cavity power reached 194 kW. However it was difficult to reach full power needed while operating on CW mode then, although beam with 90% duty cycle was transmitted through the RFQ successfully.

The pulsed proton beam with peak current of 10mA was shooting through to verify the RFQ performance. The output energy of 3.19 MeV was detected by two downstream FCTs with fine measurement of the distance in between with 0.05mm alignment error. Long pulse beam conditioning was started with beam duty factor of 50% up to 90% with 300 kW RF power feeding in the cavity. The beam transmission reached 90% with 90% beam duty factor in Sep. 25[th] of 2014 and the proton beam lasted for few minutes before being interlocked because of reaching the dump target temperature threshold. However the 60% duty factor beam passing through the RFQ operated stably for one hour, before being stopped artificially to save the dump target. Details are reported in reference [10].

The transmission has also been measured for different in cavity power with short pulsed beam and lower duty factor, the maximum transmission dropped from 97% [10] to 90% as shown in figure 9 after a period of CW conditioning during March to April of 2015. Large area of surface damage of the RFQ was observed. Figure 10 shows the surface damage status pictured from #1 power coupler port of the RFQ at April of 2015. Although the damage had already caused, the performance does not drop so much for pulsed mode with small duty factor. However, it was difficult to operate the cavity with high RF duty factor. The surface damage was considered to be the problem of not pure enough copper material and more importantly, was caused mainly because of the lacking of reflection power protection during RF conditioning.

For achieving CW operation, a new RFQ with better material was fabricated and installed in the tunnel last August replacing the old one. Noteworthy, while dissembling the old RFQ, the springs of some plug tuners were found to be burned out heavily. This should be the main contribution to the old RFQ transmission drop. The same story was happened before for the spring of the 4[th] RFQ power coupler. Figure 11 shows the burned 4# FPC RF contact spring after disassembly on September of 2014. It was thought to be the problem of not tight enough connections causing RF contact spring discharging while feeding high power into the cavity. After properly re-installing, the problem was solved for that coupler. Although the problem was solved then, the spring is indeed considered to be a weak point and also not an advisable choice for the high power CW machine. For the new design of the RFQ power coupler, the spring design had been gotten rid of for the new fabricated RFQ.

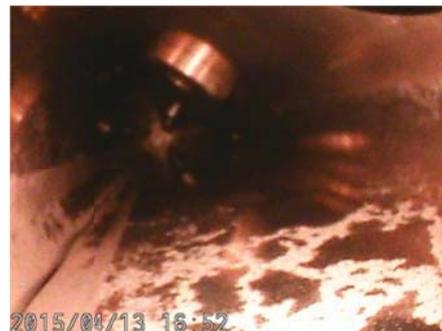

FIG. 10: RFQ surface damage status pictured from the #1 power coupler port.

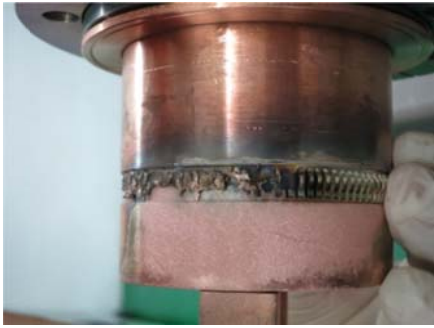

FIG. 11: The burned 4# FPC RF contact spring after disassembly [11].

However it was still kept for the plug tuners. Although the CW operation has been achieved for the new RFQ recently with around 2 mA average beam current, the spring of the plug tuners are still considered to be a problem effecting the stability operation of the RFQ.

As mentioned before, the CW commissioning of the RFQ started with small average beam current and lower beam loading effect. To keep a good quality of the beam from the source, a smaller aperture was designed to be inserted in the LEBT section for producing smaller current instead of adjusting the source itself. However the transmission scan of the RFQ was not as expected with maximum 90% transmission efficiency as shown in figure 12 versus 97% transmission for the old one with 10 mA peak current proton beam. The beam from the source is not in the centre of the aperture should be the main reason. Further beam experiment will be carried out to verify the beam transmission of the new RFQ while moving out the smaller aperture in the LEBT.

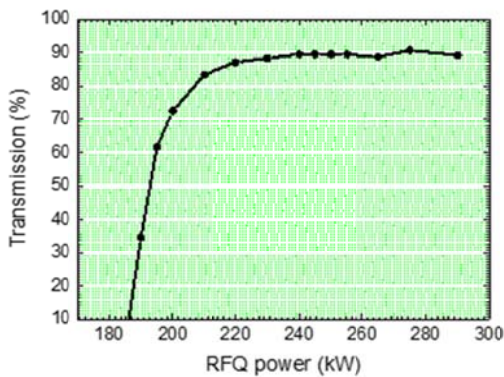

FIG. 12: Verification of tank level with transmission scan for the new fabricated RFQ with small beam current.

## C. The MEBT line

The detailed MEBT physics design is described elsewhere [12]. The MEBT is composed of six Quadruples, six steering magnets and two Bunchers. Beam diagnostic devices include six Beam Position Monitors, two Fast Current Transformers (FCT), one AC current transformer (ACCT) and three Wire Scanners. The two FCTs with 1.67m in between are used for the energy measurement. To be noted, the distance between the FCTs has to be long enough to reduce the measurement errors.

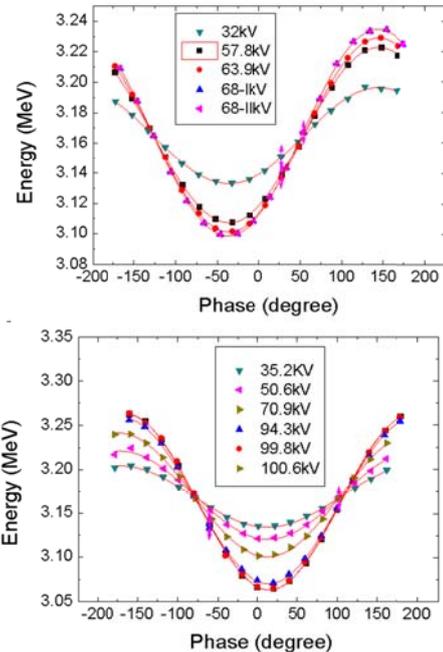

FIG. 13: The Buncher I (upper) /Buncher II (below) phase scan results.

Two downstream BPMs were used for the phase scan to determine the buncher settings. For the phase scanning, it is necessary to use the BPMs not far away for avoiding the inaccurate scanning results caused by largely expended phase spread of the beam. For the energy measurement, it is recommended to measure with buncher cavities on for the same reasons. Figure 13 show the buncher I and buncher II phase scanning results, respectively.

Different emittance measurement methods using double slits, wire scan and Quadruple scan had been tried for the RFQ exit emittance measurement at different stages. The difficulty for the double slit method is how to define the background signals, as the accuracy of the measured emittance is determined by the precision of the background signals. Multi-wires located at three different locations were used for the emittance measurement also, however a self-consistent solution could not been worked out with the existing experimental data of three existing wires. More wires are needed to give a reasonable solution or one can simply collect more data by means of Quadruple scan.

For the Quadruple scan, the first problem is the beam size fitting during data processing. Usually Gaussian curve fitting is used for the RMS beam size calculation. However, beam is usually not Gaussian shape in most of the cases especially when the Quadruple setting deviated a lot from the normal design during the gradient scans. Direct Root Mean Square formula is used in our case for the RMS beam size calculation to reduce the error from calculating method caused by unsuitable fitting formula. Detail descriptions please check in reference [13]. Another problem of Quadruple scan is the traditional emittance transfer map deducing method from the measurement position to the RFQ exit in which space charge effect is not considered. This is obviously not suitable for us while the space charge could not be neglected. The problem was solved in our case using evolutionary algorithm for finding the optimal solution for multiple objectives by calling the multi-particle tracking code TraceWin [14]. Detail descriptions is reported in reference [15]. Table III listed the emittance measurement results using Quadruple scan with space charge effect, the simulation results are also listed as a comparing.

Table III: Twiss parameters comparison between simulation and measurement results for the RFQ exit and CM1 exit.

| Parameters | $\alpha_x/\alpha_y$ | $\beta_x/\beta_y$ (mm/mrad) | $E_{n,rms,x/y}$ ($\pi$ mm.mrad) |
|---|---|---|---|
| **RFQ exit** | | | |
| Simulation | -1.3/1.46 | 0.12/0.13 | 0.21/0.2 |
| Quad. scan | -1.8/0.72 | 0.17/0.09 | 0.16/0.21 |
| **CM1 exit** | | | |
| Simulation* | -1.44/-1.75 | 1.18/1.53 | 0.22/0.21 |
| Quad. scan | -2.12/-1.97 | 1.56/1.81 | 0.29/0.27 |

*Simulation results according to the Quad. Scan measurement results at RFQ exit with 30% longitudinal mismatch assumed.

### D. Superconducting section in the Injector-I

The SC section consists of two cryomodules with fourteen $\beta_g= 0.12$ Spoke cavities, fourteen SC solenoid and fourteen Beam Position Monitors (BPM) in total. The detailed physics design is described in reference [4]. One focusing period of Spoke012 section consists of one Spoke012 cavity, one SC solenoid and one BPM. Totally fourteen periods are assembled in two cryomoudles and seven periods in each. The first cryomodule is designed to accelerate the proton beam >5 MeV. The beam is then accelerate up to 10 MeV by the second cryomodule right after. The periodical lattice is broken at the interface of the two cryomodules as necessary space has to be kept for the cold to warm transition. The break off of the periodical lattice easily leads to mismatch during commissioning despite beam dynamics results are promising, because of unavoidable input beam mismatch and accumulated static and dynamic errors of the upstream elements. Although from beam dynamics point of view, it is better to put all the elements in one cryomodule. The beam performance is much better and less sensitive to the designed parameters. Besides, the cavity efficiency would be fully exploited as at least two cavity gradients will be sacrificed for the matching from section to section if there is a break off. Also mechanically it is feasible to put all the elements of fourteen periods in one cryostat, but in reality it is difficult for the installation, alignment and maintenance. Especially when any single one element of the complex failed, the whole cryostat might have to be pulled out. In addition to the increasing risk of maintenance, more maintenance time is also needed.

Anyhow, a short cryomodule is preferred in the SC section and the space between the cryomodules is necessary but should be kept as short as possible although it would be better to insert some diagnostic devices in between for facilitating the commissioning. Besides, a vacuum pump is also helpful at this position. For the SC section of Injector-I, the space from the last cold element of the upstream cryomodule to the first cold element of the downstream cryomodule is 570mm. If one layer of cold shielding of the SC cryostat could be adopted, this distance could be even smaller, however for the CW SC section operated at 2K, two layer of cold shielding is necessary.

To verify the cavity and solenoid performance, and evaluate the cryogenic system et al, a testing cryomodule (TCM) housing two periodical periods are installed and commissioned before the formal cryomodules.

#### 1. TCM test stand

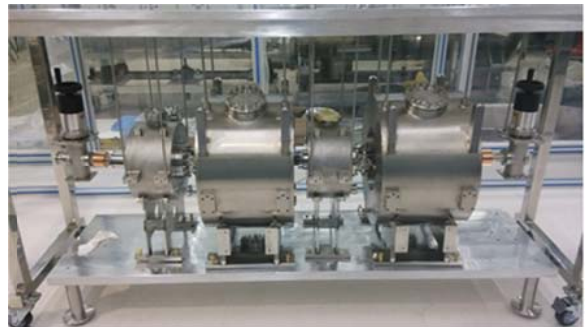

FIG. 14: The cavity string of the TCM.

The TCM turned out to be a very necessary step for technical and design validations of the formal cryomodule. We gained valuable experiences from the operation of the TCM for the improvement of hardware design including the power coupler, tuner, the LLRF controlling and beam tuning mechanism, et al. It ensured the high gradient operation of the SC cavities in the formal cryomodules.

The cavity string of the TCM housing two SC cavities, two solenoids and two BPM as shown in figure 14. It was commissioned from Jan. 26 of year 2015. The TCM was commissioned with beam duty factor of 1.5‰, the output energy achieved was 3.68 MeV with output peak current of 10.1 mA and the transmission efficiency from the entrance of RFQ to the exit of the TCM was 92%.

The maximum cavity peak gradient reached was 14MV/m with accelerating gradient of 3.1 MV/m. This was far from the design specification of 7 MV/m because of the field emission caused by cavity contamination preventing the accelerating gradient increasing further up. The cleanliness during the assembling was realized to be very important for ensuring the cavity high gradient operation. The replacement of coupler is necessarily to be carried out back in the clean room instead of in the tunnel. All of the improved technologies and methods were used for the succeeding cryomodules: CM1 and CM2. The power coupler was assembled to the cavity in the class-10 clean room instead. The power coupler structure was also changed to protect the ceramic windows. Details please check in reference [16]. The cavity operating gradient during the following commissioning stages goes much higher.

2. The first cryomodule (CM1)

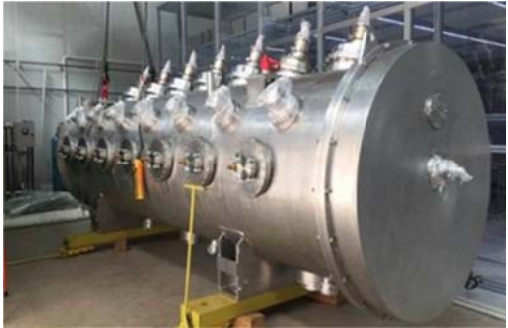

FIG. 15: The first cryomodule of Injector-I.

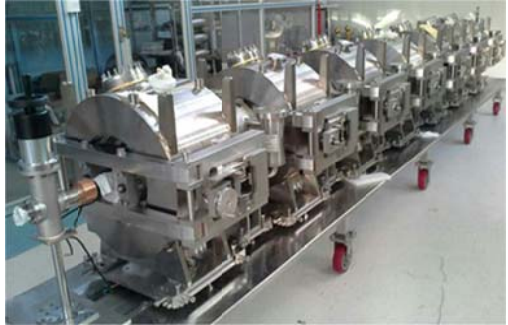

FIG. 16: The cavity string of the CM1.

The first cryomodule (CM1), as shown in figure 15, replaced the TCM after the testing cryomodule was commissioned. The CM1 include seven $\beta_g$=0.12 SC spoke cavities, seven SC solenoids and seven cold BPMs as shown in figure 16. The CM1 was conditioned beginning from May 21st of year 2015. The output energy at the exit of CM1 reached 6.05 MeV with beam peak current of 10.6 mA at Jan. 21st of year 2016 with maximum 1 ms pulsed beam and repetition frequency of 2 Hz. The maximum cavity operating gradient reached over 7 MV/m for the spoke cavities. Noteworthy, the cavity with highest operation gradient was the only one installed with piezo then.

The beam energy was measured using Time of Flight (TOF) method by two downstream FCTs. The results is shown in figure 17, the blue curve shows the signal of the 1st downstream FCT, the yellow curve shows the signal of the 2nd FCT. The beam transmission were measured by two ACCTs and one downstream DCCT. The ACCT1 is located at the entrance of the RFQ, ACCT2 is located at the entrance of the CM1 (exit of the RFQ). The DCCT is located downstream of the CM1. The beam transmission through the cryomodule is 100% while 88.4% transmission from the entrance of the RFQ to the exit of the CM1. ACCT1 current signals (yellow curve) shows in the upper graph of figure 18, DCCT current signals (purple curve) shows in the below graph of figure 18.

Figure 19 shows the twiss parameters measurement results obtained by two slits located at the exit of CM1.

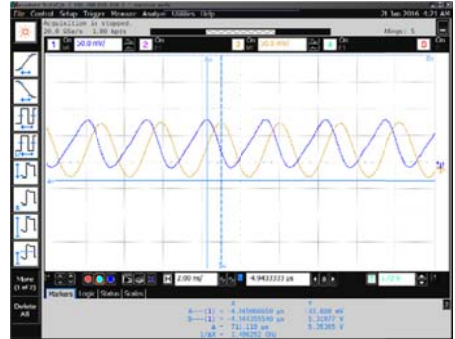

FIG. 17: The FCT3 (blue curve) and FCT4 (yellow curve) signals downstream of the CM1.

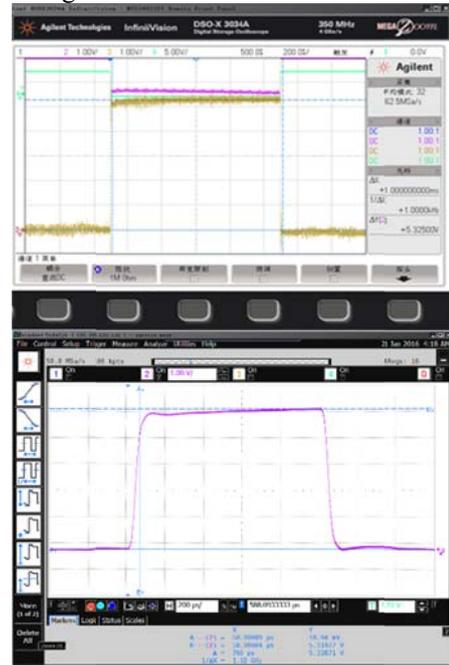

FIG. 18: ACCT signals: ACCT1 (upper graph) and DCCT (below figure).

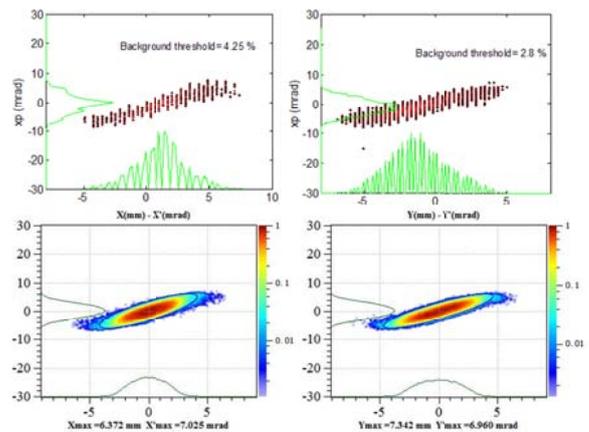

FIG. 19: Measurement emittance results by two slits at the exit of CM1 (upper graph) comparing with the simulation results (below figure).

As a comparing, the simulation results at the same position were shown in the below figure. The RFQ exit parameters used for the simulations come from the measurement results using quadrupole scan method at the MEBT section with 30% longitudinal mismatch assumed. The phase spaces are quite similar, but the twiss parameters are still not consistent with each other as shown in table II also. More research will be done to investigate the emittance and twiss parameters deviation between the simulation and measurement results.

During the design, different transverse and longitudinal phase advance ratios starting from 0.4 up to 0.75 were studied for totally fourteen periods without breaking. The maximum emittance growth percentages versus different phase advance ratios are shown in figure 20. More than 20% RMS normalized emittance growth were observed for the phase advance ratio of 0.4~0.6 with ideal input Gaussian distribution without errors included. Beam losses observed for phase advance ratio of 0.4 for weak transverse focusing with zero current periodical phase advance less than 30 degree. Finally a moderate phase advance ratio of 0.75 was chosen to avoid the envelope resonance and emittance growths. Transverse beam emittances at the exit of CM1 were measured for different phase advance ratios at Injector-I aiming to verify the conclusion obtained from the above mentioned research. However the measurements are not quite agree with the simulations. Not perfect matching in the MEBT section could be the explanation. The detail measurement results are presented in reference [16].

Table IV: The operating gradient of the fourteen SC cavities@10.67MeV

| Cav. # | 1 | 2 | 3 | 4 | 5 | 6 | 7 |
|---|---|---|---|---|---|---|---|
| $E_p$(MV/m) | 17.1 | 24.4 | 29.1 | 26.4 | 30.6 | 33.7 | 26.3 |
| Cav. # | 8 | 9 | 10 | 11 | 12 | 13 | 14 |
| $E_p$(MV/m) | 26.0 | 26.0 | 28.6 | 27.8 | 31.0 | 30.7 | 19.2 |

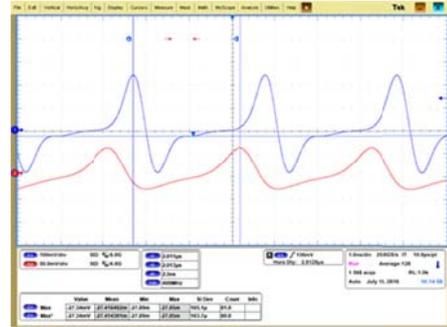

FIG. 22: The ColdBPM14 (blue curve) and TestBPM1"A" Button (red curve) signals.

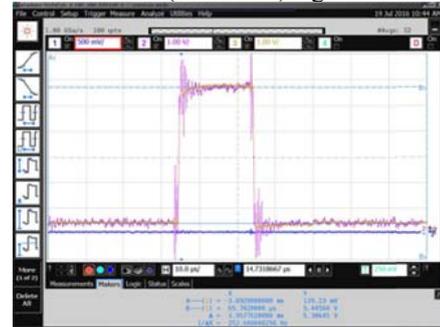

FIG. 23: ACCT2 (purple curve) and ACCT3 (yellow curve) signals.

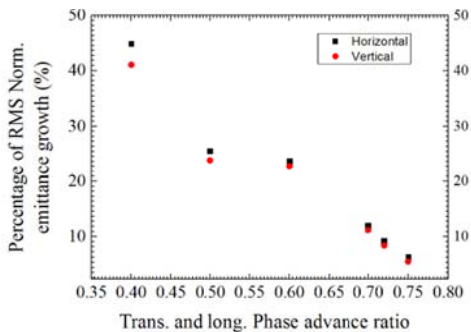

FIG. 20: Maximum emittance growth percentages versus different phase advance ratios.

3. The second cryomodule (CM2)

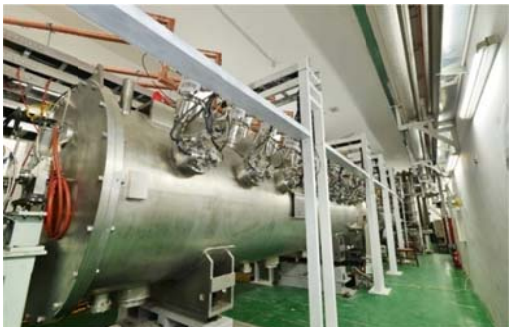

FIG. 21: The two cryomodules of Injector-I installed in the tunnel

The CM2 was installed right after the CM1. Figure 21 shows the test stand with both of the two cryomodules installed in the tunnel. The conditioning began from May 5th of year 2016 with 20 μs pulsed beam and repetition frequency of 2 Hz. The output energy at the exit of CM2 reached 10.67 MeV with beam peak current of 10.6 mA at July 19th of year 2016. The cavity operating gradients are shown in table IV. The relative low operating gradient of the first cavity originates from the contamination by an accidental vacuum leak of the upstream section.

The beam energy was measured by two BPM at the exit of the last cavity. The BPM signals shows in figure 22. The blue curve shows the signal of the last cold BPM (Cold BPM14: located right after the last cavity), the red curve shows the "A" button signal of the downstream warm BPM. The beam transmission through the cryomodule is 100% as shown in figure 23. The current was measured with two ACCT: ACCT2 (upstream of CM1: purple curve in figure 23) and ACCT3 (downstream of CM2: yellow curve in figure 23).

The beam energy divergence at the exit of the cryomodules has also been measured by the Energy Divergence Analysis (EDA) system, which include two slits, one 90 degree magnet and one faraday cup as shown in figure 1. Figure 24 shows the beam current signal (blue curve) on the faraday cup while the 90 degree magnet field and the positions of the slits were fixed. The energy divergence was measured by scanning the magnet fields. Figure 25 shows the faraday cup current signals versus magnet fields (left) and the corresponding energy versus current evolutions (right). The blue curve shows the energy divergence for the beam with lowest energy and the green curve shows the energy divergence for the beam with highest energy with beam pulse width of 20 μs and the red curve for the beam energy in the middle. The maximum energy difference for the beam with 20 μs pulse width is 0.06 MeV which is considered to be the beam loading effect at around 10mA. The RMS value of the energy divergence is 3.2‰ for 10.67 MeV and 10.6 mA peak current proton beam, which is consistent with the simulation.

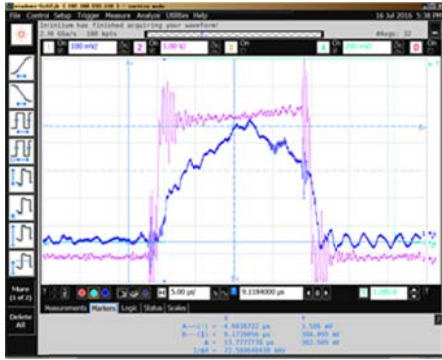

FIG. 24: Faraday cup signals of the EDA system (blue curve) and the ACCT2 current signal (purple curve).

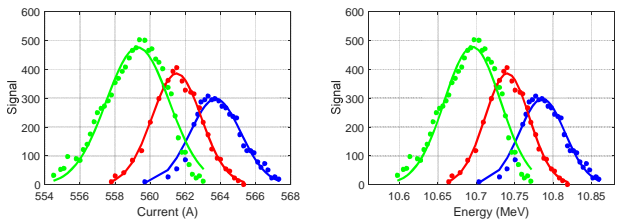

FIG. 25: The beam energy divergence at the exit of Injector-I.

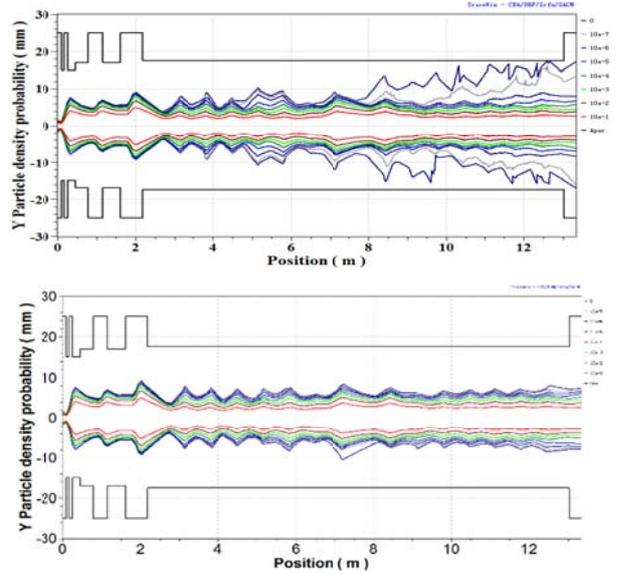

FIG. 26: The particle density probability of the Injector-I beam evolution with all the optic errors (below), above figure: with extra minus three degree offset at the entrance of the MEBT section (out of RFQ).

### E. CW commissioning of the injector-I

We accumulated valuable experiences and collected lots of data during the commissioning of injector-I with pulsed mode. However, the CW commissioning is much difficult than pulsed operation. The RFQ and SC cavity could operate on CW mode stably with full power without beam or with small duty factor beam. However, the optics in the SC section tripped frequently once the beam duty factor went higher in the beginning of the CW commissioning. Part of the SC optics trips are related closely to the beam loss. It is hard to optimize the beam transmission through the current measurement devices because the beam loss rate we concerned is beyond the accuracy of the DCCT or ACCT in used. Enough temperature sensors inside of the cryomodules might be a good solution for the beam loss localization. It could also be an assistant for the beam orbit judgement as it is difficult to do beam based alignment for the BPM with solenoids inside of the cryomodules.

Besides the transverse mismatch, the beam being matched longitudinally is also important. Simulation shows that if the beam is mismatched longitudinally at the beginning of the linac, part of the peripheral particles easily rotating out of the longitudinal bucket with optics error accumulating along the linac and finally lost in the back-end of the linac after being coupled to transverse plane. Typical beam power density evolution with optic errors coupling with bigger entrance longitudinal mismatch is shown in figure 26 (above). This figure shows the simulation results of the Injector-I beam evolution with extra minus three degree offset at the entrance of the MEBT section (out of RFQ). As shown in the above figure, the mismatch was quickly magnified after beam getting though the matching section between the two cryomodule (6 to 7 meter position in the figure) and part of the beam coupled to transverse and begin to loss on the beam pipe at the back-end of the Injector. The most serious loss happens at the very end of the linac. This is confirmed in the experiment. The Injector-I simulation results without input errors (with all the optics errors added) are shown in figure 26 (below) as a comparing.

Lots of efforts have been made to keep the stability of the machine when shooting CW beam through. The stable operation time went from dozen seconds up to several minutes and finally more than twenty minutes with output energy of 10 MeV and average beam current of ~2 mA. The maximum stable operation time reached until now is 23 min with output energy of 10 MeV and average current of 1.6 mA CW proton beam. Figure 27 shows the output energy at the exit of RFQ/cryomodule and the beam current at that period. The RFQ could deliver higher current CW beam up to 10mA, but the SC section could not handle high intensity beam while beam loading effect is not compensated and frequency control loop is absent.

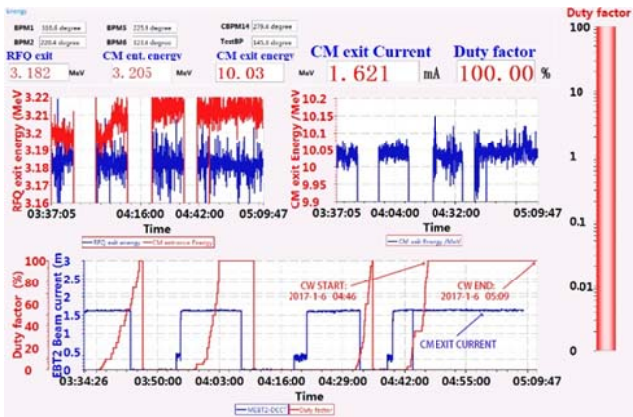

FIG. 27: CW commissioning results of Injector-I

F.  Conclusion

The China ADS injector-I testing facility has been commissioned successfully using pulsed and CW beam. The maximum energy achieved at the exit of the Injector is 10.67 MeV with peak beam current of 10.6 mA. CW proton beam with energy of 10 MeV and average beam current of around 2 mA has been obtained at the exit of the linac recently. The CW beam with average beam current of 1.62 mA lasted for 23 minutes stably without trip of any devices. Preliminary experiment results that have been obtained are encouraging but further work is still needed to do for better understanding of the phenomena that occur in high duty cycle operation of the linac. CW commissioning of the Injector is still ongoing. Beam loading effect of the SC section will be compensated and frequency control loop of the SC cavity will be added to ensure CW operation with higher average beam current and longer operation time. More experiments as well as some improvements are foreseen.